\documentclass[aps,prl,twocolumn,showpacs, groupedaddress]{revtex4}
\usepackage{amsmath}
\usepackage{amssymb}
\usepackage{epsfig}
\usepackage{multirow}
\usepackage{booktabs}
\usepackage{dcolumn}

\begin{document}
\preprint{}
\title{Quantum simulation of a system with competing two- and three-body interactions}
\thanks{xhpeng@ustc.edu.cn, djf@ustc.edu.cn, Dieter.Suter@tu-dortmund.de. J. Zhang is presently at Institute for Quantum Computing, University of Waterloo, Canada.}

\author{Xinhua Peng$^{1,2}$}
\author{Jingfu Zhang$^{2}$}
\author{Jiangfeng Du$^{1}$}
\author{Dieter Suter$^{2}$}

\affiliation{$^{1}$Hefei National Laboratory for Physical Sciences at Microscale and 
Department of Modern Physics, University of Science and Technology of China, Hefei, Anhui 230026, People's Republic of China}
\affiliation{$^{2}$Fakult\"{a}t Physik, Technische Universit\"{a}t Dortmund, 44221 Dortmund, Germany}

\date{\today}

\begin{abstract}
Quantum phase transitions occur at zero temperature, when the ground state of a Hamiltonian undergoes a qualitative change
as a function of a control parameter. 
We consider a particularly interesting system with competing one-, two- and three-body interactions.
Depending on the relative strength of these interactions, the ground state of the system can be a product state,
or it can exhibit genuine tripartite entanglement.
We experimentally simulate such a system in an NMR quantum simulator and
observe the different ground states.
By adiabatically changing the strength of one coupling constant, we push the system from one ground state
to a qualitatively different ground state.
We show that these ground states can be distinguished and the transitions between them observed 
by measuring correlations between the spins or the expectation values of suitable entanglement witnesses.
\end{abstract}

\pacs{03.67.-a, 03.65.Ud, 03.67.Mn}

\maketitle

\emph{Introduction.--} 
At zero temperature, a system can undergo a quantum phase transition (QPT) to a new ground state 
as a result of a change in a parameter of the Hamiltonian \cite{SachdevBook:1999aa}. 
Well-known examples are the superconductor-insulator transition and the paramagnetic-antiferromagnetic transition in quantum magnets. 
QPTs were experimentally observed in magnetic systems \cite{Ronnow:2005aa}, heavy-fermion metals \cite{Custers:2003aa}, 
common metals \cite{Yeh:2002aa} and Bose-Einstein condensation \cite{Greiner:2002aa}.
The investigation of QPTs is useful for discovering novel materials \cite{Canfield:2008aa}, 
and some QPTs show interesting entanglement characteristics \cite{Osterloh:2002aa}. 
A collection of reviews \cite{Canfield:2008aa,QPTfocus} on the topic of QPTs reported the current status and recent developments in this field.

In most systems studied, attention was focused on two-body interactions, which are most readily accessible experimentally. 
On the other hand, systems with three-body interactions have been shown to exhibit exotic quantum phases 
in their ground states \cite{Buchler:2007aa}, 
such as topological phases or spin liquids and a chiral phase.  
However, it is difficult to observe these properties in experiments. 
The main challenges are: 
($i$) to identify experimentally accessible systems with three-body interactions; 
($ii$) to experimentally control the variation of the system Hamiltonian in a sufficiently precise manner;
($iii$) to characterize the resulting ground state. 

QPTs are generally associated with strongly correlated quantum systems, 
which cannot be efficiently simulated on classical computers because the required computational resources 
grow exponentially with the system size. 
As suggested by Feynman \cite{Feynman:1982aa} and proved by Lloyd \cite{5690}, however, 
a quantum computer can efficiently perform this kind of simulations 
and provide new insight into strongly correlated quantum systems, including QPTs. 
Previous examples of such studies include the simulation of a three-body Hamiltonian 
\cite{Tseng:1999aa} and quantum magnets in a two-spin Heisenberg chain 
using nuclear spins \cite{Peng:2005aa} as well as trapped ions \cite{Friedenauer:2008aa}. 

In this Letter, we experimentally simulate the smallest spin system involving a three-body interaction
in an NMR quantum simulator. 
Such systems are the building blocks of triangular Ising nets \cite{Wannier:1950} and triangular spin ladders \cite{Pachos:2004ab}, 
which have been studied in detail in the context of quantum statistics \cite{Francesco:1997wj} and condensed matter theory \cite{Buchler:2007aa, Ong:2004fe}. 
In general, the competition between the different interactions results in QPTs. 
Here we prepare the system in the ground state of the Hamiltonian and drive it from
one phase into a different phase by adiabatically changing the Hamiltonian.
By quantifying different types of entanglement,
or by using suitable entanglement witnesses \cite{PhysRevA.63.050301}, 
we successfully detect the quantum transition induced by the three-body interactions.
In the thermodynamic limit, this transition should correspond to a novel type of QPT \cite{Pachos:2004ab}.

\emph{System.--} 
Consider a chain of $N$ spins 1/2 in a uniform magnetic field, interacting
by Ising-type nearest-neighbor two-body and three-body couplings: 
\begin{eqnarray}
\mathcal{H} & = & \omega_z \sum \sigma _{z}^{i}  + \omega_x \sum \sigma _{x}^{i}   \nonumber \\
  & &  + J_2  \sum \sigma _{z}^{i} \sigma _{z}^{i+1} + J_3 \sum \sigma _{z}^{i}\sigma _{z}^{i+1} \sigma _{z}^{i+2},
\label{e.H}
\end{eqnarray}
where the $\sigma^i_z$ are the Pauli operators, 
$\omega _{z}$ and $\omega _{x}$ the strengths of the longitudinal and transverse magnetic fields, 
and $J_2$, $J_3$ the two-body and three-body coupling constants. 
The competition between the three kinds of interactions (one-, two- and three-body)
determines the ground state of the system. 
We first consider two limiting cases, corresponding to the two- and three-spin Ising models.
In both cases, we discuss explicitly the situation for positive coupling constants $\omega_z,\omega_x, J_2, J_3 > 0$.

(\textit{i}) Two-spin Ising model ($J_3 = 0$) \cite{SachdevBook:1999aa}: (\textit{ia}) in a longitudinal magnetic field ($\omega_x$=0). 
The level crossing at $J_2 = \omega_z$ corresponds to a first-order phase transition of the ground state 
from a paramagnetic state $\vert \psi_g ^{P} \rangle = \vert .. \downarrow \downarrow .. \rangle $ 
in the weak-coupling case ($J_2  < \omega_z $) to a two-fold degenerate, antiferromagnetically ordered ground state: 
$\vert \psi_g ^{AF} \rangle = \{ \vert .. \uparrow \downarrow .. \rangle, \vert ..\downarrow \uparrow.. \rangle \}$ in the strong-coupling case ($J_2 >\omega_z$ ). 
In the case of ferromagnetic coupling ($J_2 <0$), no QPT occurs.
(\textit{ib}) Two-spin Ising model in a transverse magnetic field (i.e., $\omega_z$=0). 
This well-studied model exhibits a second-order QPT at $J_2 = \omega_x$ in the thermodynamic limit \cite{SachdevBook:1999aa}, 
where the ground state changes from the paramagnetic phase 
$\vert \psi_g ^{P'} \rangle = \vert .. \leftarrow \leftarrow .. \rangle$ 
to the antiferromagnetically ordered, doubly degenerate ground state $\vert \psi_g ^{AF} \rangle$. 
Here $\vert \leftarrow \rangle = (\vert \uparrow \rangle - \vert \downarrow \rangle) / \sqrt{2}$. These two cases are exactly solvable and the induced phase transitions can be detected by the traditional two-point correlation functions \cite{SachdevBook:1999aa}. 

(\textit{ii}) Three-spin Ising model with $J_2 = 0$ in a transverse magnetic fields (i.e., $\omega_z$=0):  
This model is not exactly solvable, but numerical simulations predict a critical point at $J_3 = \omega_x$.
The strong-coupling phase ($J_3 > \omega_x$) is fourfold degenerate:
$\vert \psi_g ^{F} \rangle = \{ \vert .. \downarrow \downarrow \downarrow . \rangle 
, \vert .. \downarrow \uparrow \uparrow .. \rangle, \vert .. \uparrow \downarrow \uparrow ..\rangle, 
\vert .. \uparrow \uparrow \downarrow .. \rangle\}$, 
while the non-degnerate ground state of the weakly-coupled phase is the paramagnetic state
$\vert \psi_g ^{P'} \rangle$ \cite{PhysRevB.26.6334}. 

\begin{figure}[tb]
\begin{center}
\includegraphics[width = 0.99\columnwidth]{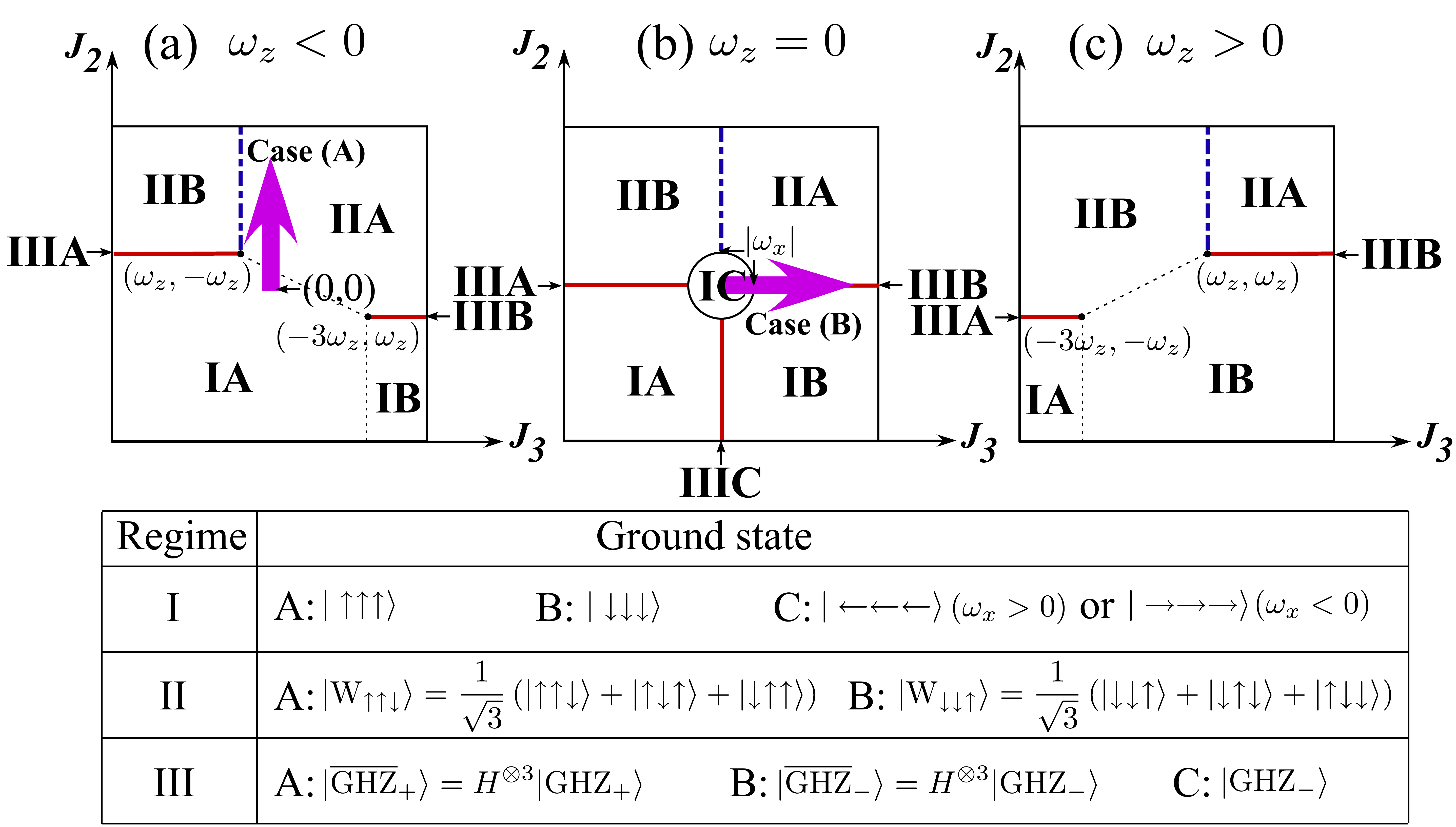}
\end{center}
\caption{(Color Online) Schematic phase diagram of the ground state of the Hamiltonian $\mathcal{H}$ for $N=3$. 
Here $\vert \mathrm{GHZ}_{\pm} \rangle 
= (\vert \uparrow \uparrow \uparrow \rangle \pm \vert \downarrow \downarrow \downarrow \rangle)/ \sqrt{2}$ and $H$ is the Hardamard gate. 
The arrows in (a) and (b) represent the adiabatic evolutions discussed in the experimental section. 
For (a) and (c), $\omega_x$ is assumed to be small, i.e., $\vert \omega_x \vert \ll \vert \omega_z \vert$.}
\label{phasediag}
\end{figure}

In the following, we demonstrate these features in a proof-of-principle experiment using $N=3$ spins and periodic boundary conditions 
$\sigma_z^{N+1} = \sigma_z^{1}$.
Figure \ref{phasediag} shows the ground state of the system.
The competition between the three kinds of interactions (one-, two- and three-body)
results in different ground state phases (product-, W-, and GHZ-states). 
In regime I, the ground states are product states, in phase II, they are W-type states,
and in regime III, they are GHZ-type states.
The GHZ and W states are the only two inequivalent kinds of genuine tripartite entanglement in a 3-spin system \cite{Dur:2000aa}. As a result, such a simple model allows us to experimentally study the two inequivalent states of genuine tripartite entanglement and the novel phenomena induced by three-spin interactions. 

\begin{figure}[htb]
\begin{center}
\includegraphics[width= 0.99\columnwidth]{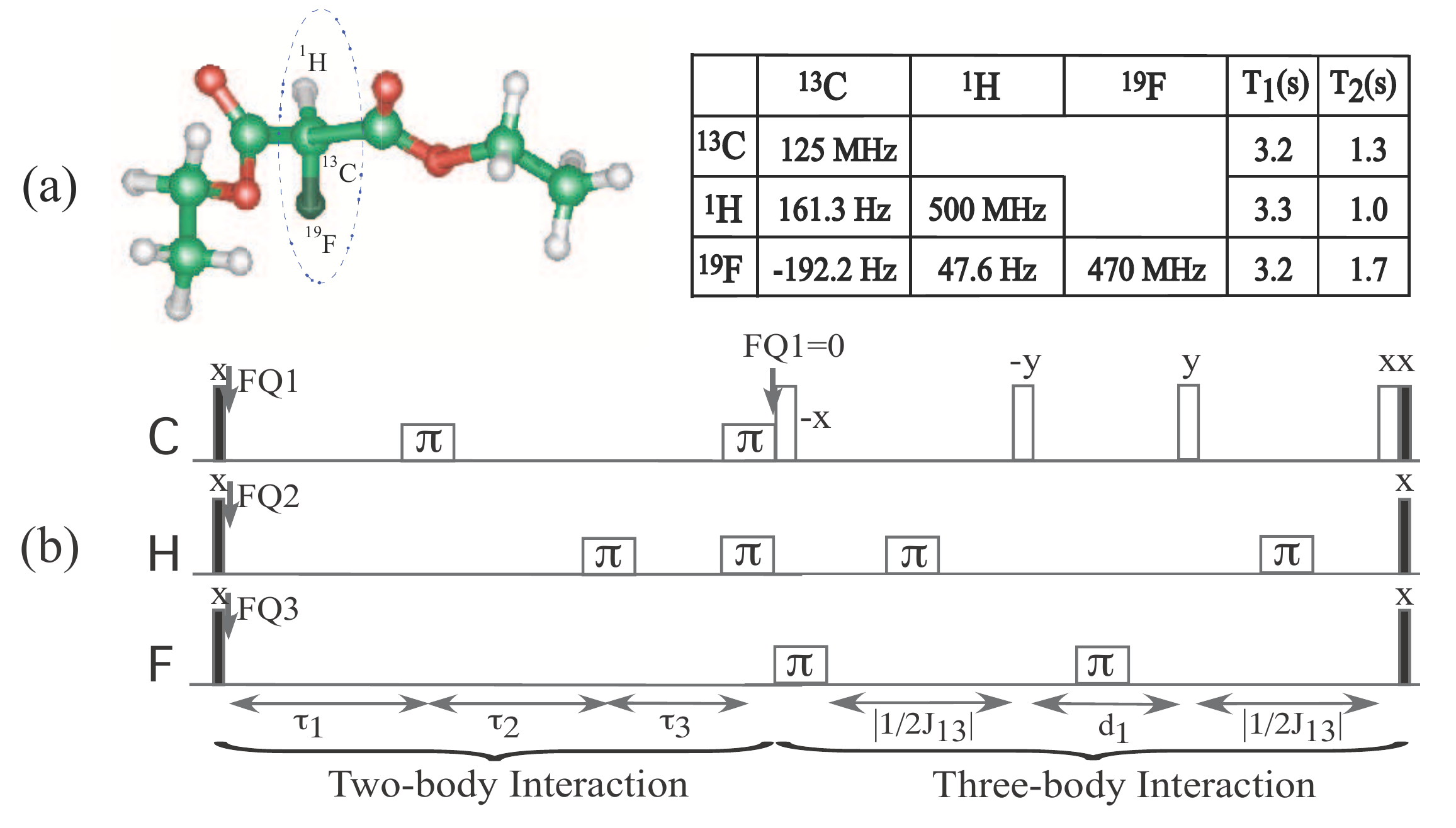}
\end{center}
\caption{(Color Online) (a) Molecular structure and properties of the quantum register: Diethyl-fluoromalonate. The oval marks the three spins used in the experiment. 
The table on the right summarizes the relevant NMR parameters  measured at room temperature on a Bruker Avance II 500 MHz (11.7 Tesla) spectrometer, 
i.e., the Larmor frequencies $\omega_i/2 \pi$ (on the diagonal), the J-coupling constants $J_{ij}$ (below the diagonal), 
and the relaxation times T$_1$ and T$_2$ in the last two columns. 
(b) Pulse sequence for simulating the Hamiltonian of Eq. (\ref{e.H}). 
The narrow black rectangles represent small-angle rotations, 
the narrow empty rectangles denote 90$^{\circ}$ rotations and the wide ones denote the refocusing 180$^{\circ}$ pulses. 
The delays are $\tau_i = J_2 \tau /[1/ (\pi J_{ij})+1 / (\pi J_{jk})]$ with $(i, j, k)$ an even permutation of $(1,2,3)$ and $d_1 = 2 J_3 \tau/(\pi J_{12})$. 
The offsets between the irradiation frequencies and the corresponding Larmor frequencies are 
$FQ1= 2\omega_z \tau / (\tau_1-\tau_2+ 3\tau_3)$, $FQ2 = 2 \omega_z \tau / (\tau_1+\tau_2-\tau_3)$ 
and $FQ3 = 2 \omega_z \tau / (\tau_1+\tau_2+\tau_3)$.}
\label{molecule}
\end{figure}

\emph{Quantum simulation.--}
We chose the Diethyl-fluoromalonate molecule as a three-spin NMR quantum simulator, in which the $^{13}$C, $^{1}$H, and $^{19}$F nuclear spins are represented by spins 1, 2, and 3, respectively. 
The molecular structure and the relevant parameters are shown in Fig. \ref{molecule} (a).
The natural Hamiltonian of the system is 
$\mathcal{H}_{\mathit{NMR}}=  \sum_{i = 1}^{3} \frac{\omega_i}{2} \sigma^i_z + \sum_{i<j,i=1}^{3} \frac{\pi J_{ij}}{2} \sigma^i_z\sigma^j_z $. 

From this natural Hamiltonian, we generate the model Hamiltonian \eqref{e.H} 
with equal two-body coupling strengths by a suitable refocusing scheme \cite{Linden:1999ab}. 
The three-body interaction can be simulated by a combination of two-body interactions and RF pulses \cite{Tseng:1999aa}.
Since all terms in the Hamiltonian  \eqref{e.H}, except the transverse field term, commute with each other, 
we expand the overall evolution by the following concatenation:
$e^{-i\mathcal{H}\tau } = e^{-i\mathcal{H}_{x}\tau /2}e^{-i\mathcal{H}_{z}\tau }e^{-i\mathcal{H}_{x}\tau /2} + O(\tau^3)$.
Here, $\mathcal{H}_{x}=\omega_x \sum \sigma _{x}^{i} $ and $\mathcal{H}_{z}=\mathcal{H}-\mathcal{H}_{x}$. 
This expansion faithfully represents the targeted evolution provided the duration $\tau$ is kept sufficiently short. 
Fig. \ref{molecule}(b) shows the pulse sequence that realizes $e^{-i\mathcal{H}\tau }$.

In order to observe the system undergoing the transitions, it must always be 
close to the instantaneous ground state of the time-dependent Hamiltonian. 
This was achieved by quantum adiabatic evolution \cite{Messiah:1976aa}, which requires: 
$(i)$ the system starts in the ground state of $\mathcal{H}(0)$, $(ii)$ $\mathcal{H}(t)$ 
changes sufficiently slowly to satisfy the adiabatic condition.
For the experimental implementation, we discretized the time-dependent Hamiltonian $ \mathcal{H}(t)$ into $M+1$ segments $
\mathcal{H}(m) = \mathcal{H}[C(\frac{m}{M}T)]$ with $m = 0, 1,...,M$ \cite{Steffen:2003aa}, 
where $T$ is the total duration of the adiabatic passage and $C$ is the control parameter in $ \mathcal{H}(t)$. 
The adiabatic condition is satisfied when both $T, M \to \infty$ and the duration of each step $\tau \to 0$.

To observe the different types of transitions in the Hamiltonian (\ref{e.H}), we chose two different parameter sets: 
\begin{description}
\item[Case A:] Transition from a product state to a W-type entangled state: 
the control parameter is the two-body coupling strength $J_2$,
which varies from $J_2(0) = 0$ to $J_2(T)= 2$.
The other parameters of the Hamiltonian are constant, $\omega_z = -2$, $\omega_x = 0.09$ and $J_3 = 0$, 
as shown in Fig. \ref{phasediag} (a).
\item[Case B:] Transition from a product state to a GHZ-type entangled state: 
the control parameter is the three-body coupling strength $J_3$, 
which varies from $J_3(0) = 0$ to $J_3(T)= 2$, with $\omega_z =J_2 =0$ and $\omega_x = 0.12$, 
as shown in Fig. \ref{phasediag} (b). 
\end{description}
We used a hyperbolic sine for the time dependence of the control parameters $J_2(t)$ and $J_3(t)$ 
in the experiments \cite{Peng:2005aa}.  

\begin{figure}[htb]
\begin{center}
\includegraphics[width= 0.99\columnwidth]{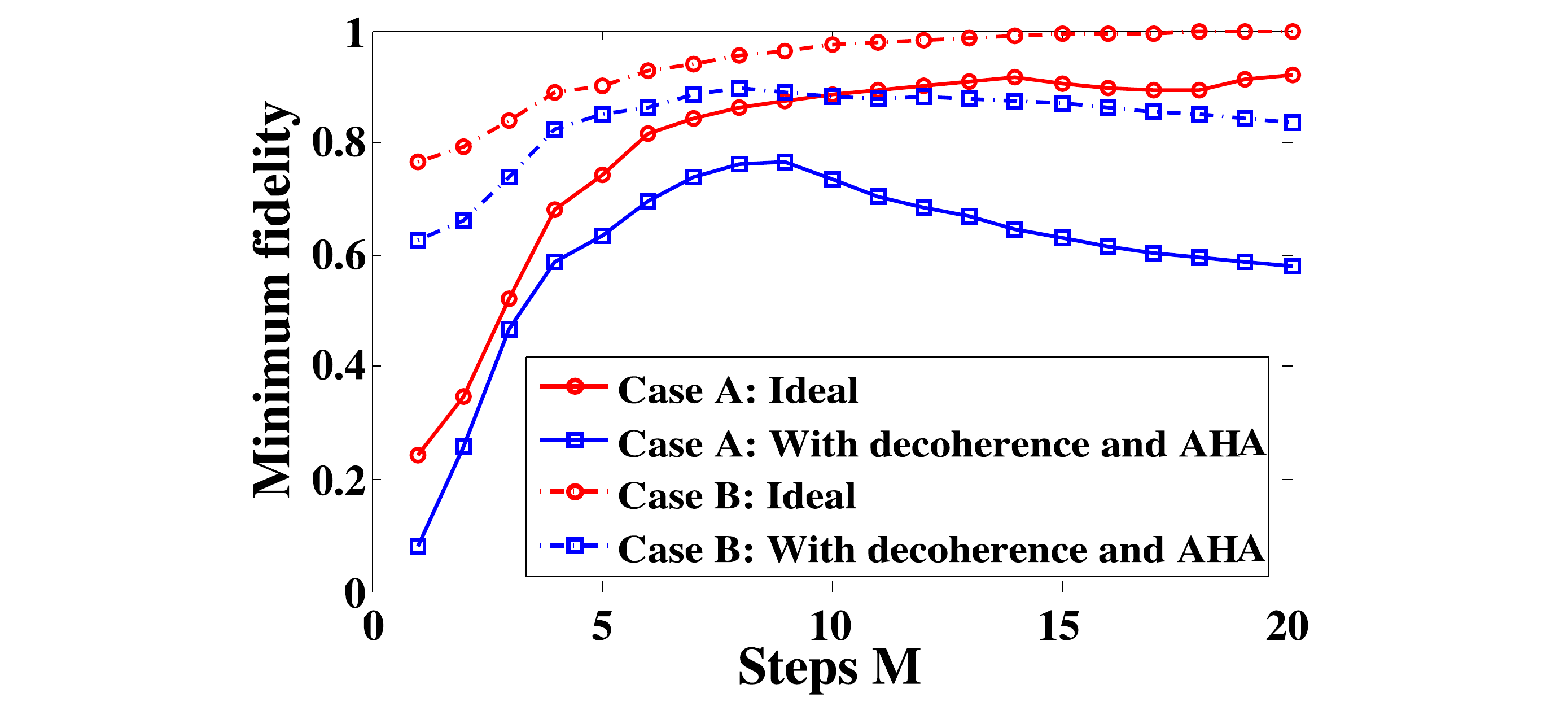}
\end{center}
\caption{(Color Online) Numerical simulation of the minimum fidelities during the adiabatic passage 
vs. the number of steps for the ideal (piecewise constant) Hamiltonian ($\bigcirc$) and for the more realistic
case that includes the actual pulse sequence and the effect of decoherence ($\Box$) 
for both cases A and B. }
\label{F_M}
\end{figure}

To determine the optimal number $M$ of steps in the adiabatic transfer, 
we used a numerical simulation of the minimum fidelity encountered during the scan 
as a function of the number of steps into which the evolution is divided (see Fig. \ref {F_M}). 
The fidelity is calculated as the overlap of the state with the ground state at the relevant position. 
Using the decoherence model of Vandersypen \textit{et al.} \cite{Vandersypen:2001aa}, 
we also simulated the effects of decoherence under the actual (piecewise constant) Hamiltonian
generated by the pulse sequence of Fig. \ref{molecule} (b).
The resulting fidelity of these points, which is represented by the boxes
in Fig. \ref{F_M}, is consistently lower than that of the ideal Hamiltonian
and reaches a maximum for $M\approx 8$.
For the experiments, we therefore chose $M = 8$. 

%

\begin{figure}[htb]
\begin{center}
\includegraphics[width = 0.99\columnwidth]{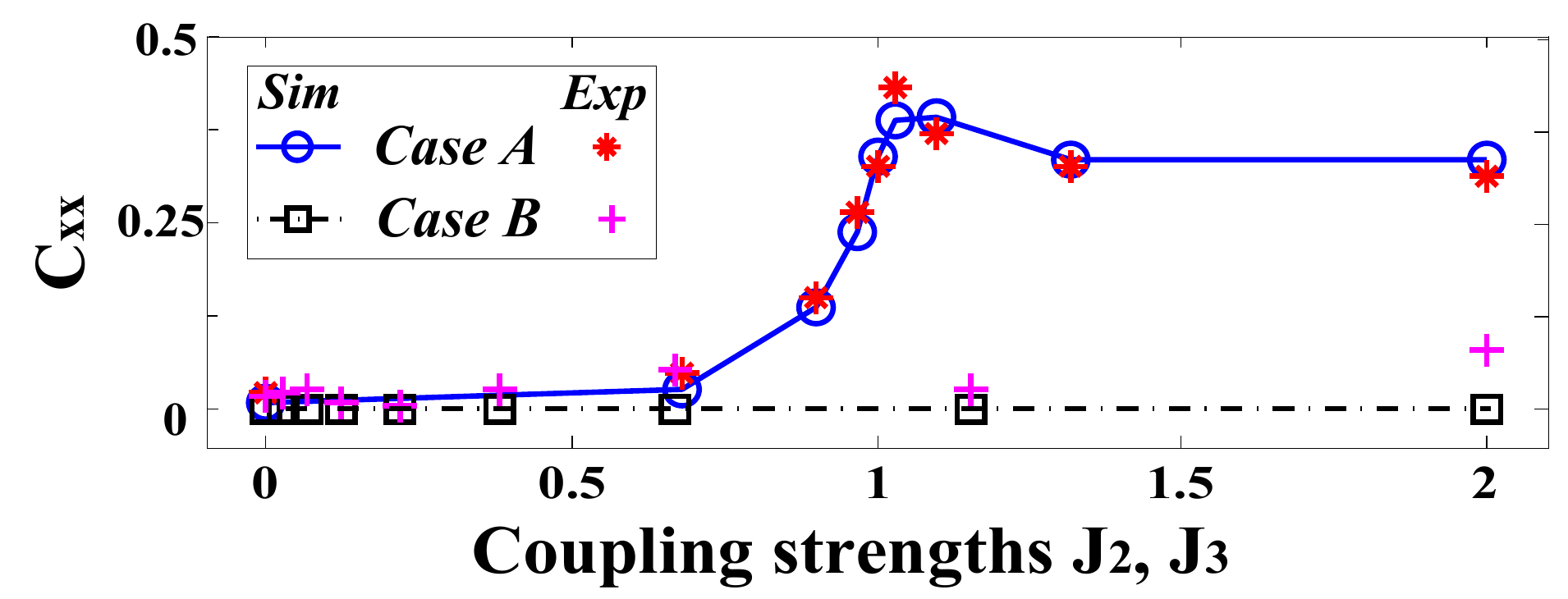}
\end{center}
\caption{(Color Online) Experimental detection of quantum transition points by two-spin correlations $C_{xx}^{exp}$ ($\ast$ and $+$), 
together with the simulated correlations $C_{xx}^{sim}$ ($\bigcirc$ and $\Box$), which take the effect of decoherence into account. 
The effective decoherence time $T_2^{eff}$ was estimated as $150\,$ms for Case A and $600$\,ms for Case B.}
\label{Cxx}
\end{figure}

\emph{Detection of transition points.--} 
The conventional approach to detect the system undergoing an entangling phase transition 
is to measure two-spin correlations like 
$C_{xx} = \frac{1}{3} \sum_{i \ne j} \langle \sigma_x^{i} \sigma_x^{j} \rangle .$
The result of such a measurement is shown in Fig.\,\ref{Cxx}.
The result clearly shows the expected transition for Case A:
The two-spin correlation goes through a step-like increase at $J_2 \approx 1$,
consistent with the expectation that the ground state should change from a product state to a W state
at this value of the couplings strength.
However, in Case B, where the system should start in a product state and end up in a GHZ state, 
the measured two-spin correlations give no indication of this transition.

In order to obtain a clean signature of this novel transition, induced by three-body interactions,
we employed entanglement witnesses \cite{PhysRevA.63.050301}.
These witness operators can always be used to detect various forms of multipartite entanglement, 
provided we have some \textit{a priori} knowledge about the states under investigation. 
In our context, the two witness operators
$
\mathcal{W}_{\mathrm{W}} = \frac{2}{3} \mathbf{1} - \vert \mathrm{W}_{001} \rangle \langle \mathrm{W}_{001} \vert
$
and
$
\mathcal{W}_{\mathrm{GHZ}} = \frac{3}{4} \mathbf{1} - \vert \mathrm{GHZ}_- \rangle \langle \mathrm{GHZ}_- \vert
$
are useful for detecting multipartite entanglement.
If their expectation value is negative, they indicate the presence of genuine tripartite entanglement; 
$Tr(\rho \, \mathcal{W}_{\mathrm{GHZ}})<0$ specifically detects GHZ-type entanglement.
To perform these measurements, we applied a basis transformation to the density operator and destroyed the off-diagonal elements
with pulsed magnetic field gradients to implement the projective measurement \cite{Cory:1998ab}.
The populations were read out by applying, in three separate experiments,  $\pi/2$ readout pulses to each qubit
and measuring the resulting transverse magnetization.

We first initialized the system into the ground state of the initial Hamiltonians \cite{Cory:1998aa} and then varied the 
Hamiltonians adiabatically. 
To minimize nonadiabatic effects, the scan had to be relatively slow, lasting about $146$ ms for Case A and $62$ ms for Case B. 
Spin relaxation therefore reduced the signal for the later times compared to the initial part of the scan.
To separate this effect from the effect of the transition and obtain a cleaner signature, we first determined
the overall signal decay and rescaled the data to remove this effect.

\begin{figure}[htb]
\begin{center}
\includegraphics[width = 0.99\columnwidth]{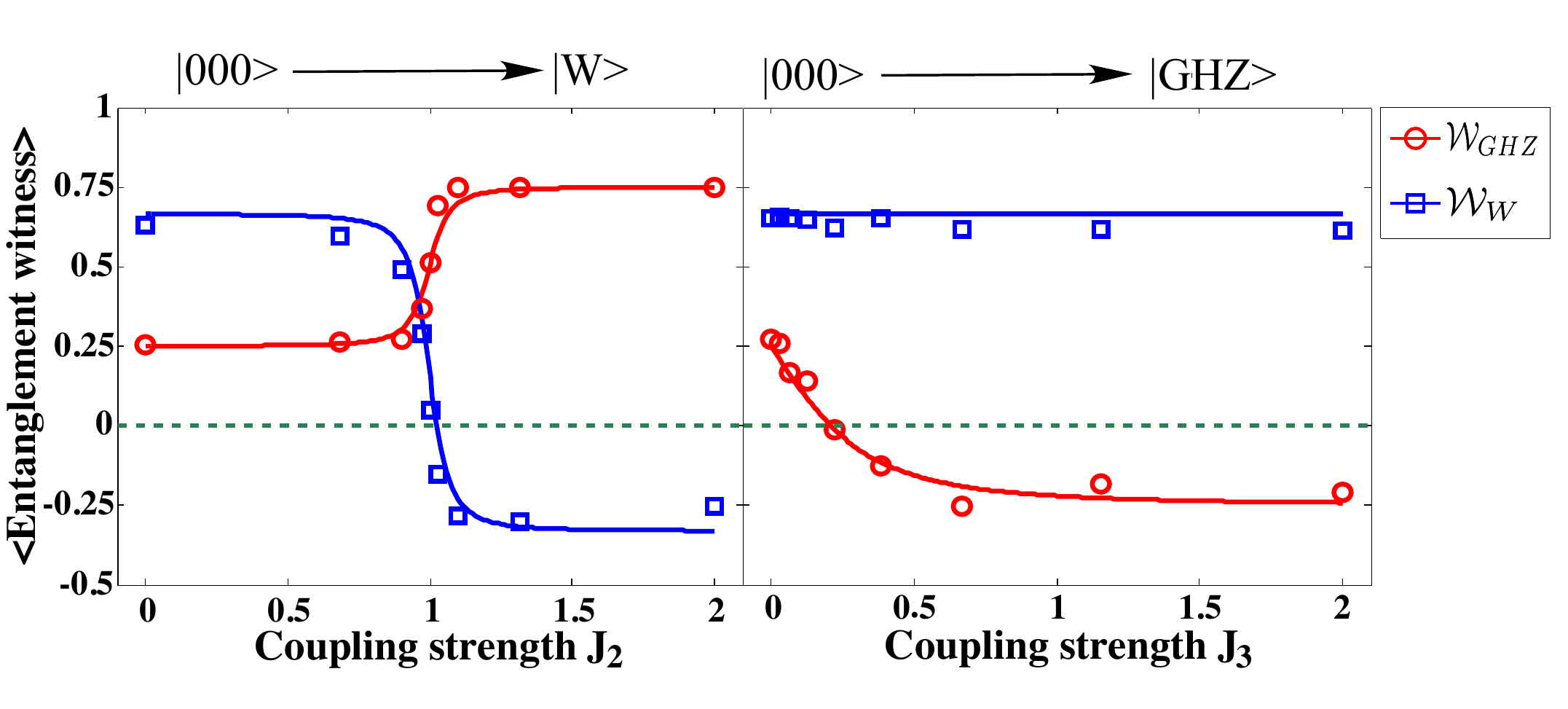}
\end{center}
\caption{(Color Online) Experimental detection of quantum transition points by entanglement witnesses for Case A (left hand side)  
and Case B (right hand side). 
The solid lines represent the theoretical expectations. }
\label{ExpResult}
\end{figure}

Fig. \ref{ExpResult} summarizes the experimental results for both cases, using both entanglement witnesses.
Both transitions are now clearly visible, and the tripartite entanglement is present in both final states. 
Together with the measurement of the two-spin correlations in Fig. \ref{Cxx}, we can also differentiate between the two types of transitions.
In Case A, the increasing strength of the two-body interaction $J_2$ pushes the system 
from a product state to a W state, 
but in Case B, the increase in the three-body interaction strength $J_3$ generates a GHZ state.

We also determined the details of the final state at the end of the scan, 
by performing complete quantum state tomography \cite{Chuang:1998aa}. 
The experimentally generated state $\rho_{exp}$ had an overlap of $F = \vert  \langle \psi_{id} \vert\rho_{exp}\vert \psi_{id} \rangle \vert$ 
with the ideal state $\vert \psi_{id} \rangle$ of 0.61 for the W state and 0.73 for the GHZ state. 
The reduced fidelity mainly results from relaxation during the long adiabatic passage. 
As in the case of the witness measurements during the scan, 
we also removed the effect of spin relaxation from the measured final states
by calculating the `experimental fidelity' \cite{Fortunato:2002aa} 
$F = \vert  \langle \psi_{id} \vert\rho_{exp}\vert \psi_{id} \rangle / Tr(\rho_{exp}^2) \vert$.
The observed data yielded 
$F(\rho^{W}_{exp}) = 0.90$ and $F(\rho^{\mathrm{GHZ}}_{exp}) = 0.92$, in excellent agreement with theoretical expectations.

\emph{Conclusion.--} 
Using an NMR quantum simulator, we have performed an experimental quantum simulation of 
a system with competing one-, two- and three-body interactions.
By adiabatically changing the Hamiltonian, we could observe the system
undergoing two different entangling transitions. 
The generation of entanglement confirms that the transitions are of quantum-mechanical origin (analogous to quantum fluctuations in QPTs). 
In particular, with suitable entanglement witnesses as observables, we detected a novel transition induced by three-body interactions that is qualitatively different from the states that can be characterized by two-spin correlations. 

We consider our present experiment based on a universal simulator as a first experimental step that demonstrates how one can obtain a clear picture of phase transitions in ground states of many-body Hamiltonians, with the help of tools coming from the theory of entanglement. Systems consisting of more than three subsystems can display even more different phases in their ground states. 
We expect that entanglement witnesses will turn out to be useful for detecting ground-state phases in larger systems,
together with other entanglement measures (e.g., the global entanglement \cite{oliveira:170401}).

\begin{acknowledgments}
 We thank Y. J. Deng, D. Bruss and H. Kampermann for helpful discussions. This work is supported by the CAS, NNSFC, and the DFG through Su 192/19-1.
\end{acknowledgments}


\begin{thebibliography}{47}
\expandafter\ifx\csname natexlab\endcsname\relax\def\natexlab#1{#1}\fi
\expandafter\ifx\csname bibnamefont\endcsname\relax
  \def\bibnamefont#1{#1}\fi
\expandafter\ifx\csname bibfnamefont\endcsname\relax
  \def\bibfnamefont#1{#1}\fi
\expandafter\ifx\csname citenamefont\endcsname\relax
  \def\citenamefont#1{#1}\fi
\expandafter\ifx\csname url\endcsname\relax
  \def\url#1{\texttt{#1}}\fi
\expandafter\ifx\csname urlprefix\endcsname\relax\def\urlprefix{URL }\fi
\providecommand{\bibinfo}[2]{#2}
\providecommand{\eprint}[2][]{\url{#2}}

\bibitem[{\citenamefont{Sachdev}(1999)}]{SachdevBook:1999aa}
\bibinfo{author}{\bibfnamefont{S.}~\bibnamefont{Sachdev}},
  \emph{\bibinfo{title}{Quantum Phase Transition}}
  (\bibinfo{publisher}{Cambridge University Press},
  \bibinfo{address}{Cambrige}, \bibinfo{year}{1999}).


\bibitem[{\citenamefont{R{\o}nnow et~al.}(2005)\citenamefont{R{\o}nnow,
  Parthasarathy, Jensen, Aeppli, Rosenbaum, and McMorrow}}]{Ronnow:2005aa}
\bibinfo{author}{\bibfnamefont{H.~M.} \bibnamefont{R{\o}nnow}} \bibnamefont{\textit{ et~al.}}, \bibinfo{journal}{Science}
  \textbf{\bibinfo{volume}{308}}, \bibinfo{pages}{389} (\bibinfo{year}{2005}).

\bibitem[{\citenamefont{Custers et~al.}(2003)\citenamefont{Custers, Gegenwart,
  Wilhelm, Neumaier, Tokiwa, Trovarelli, Geibel, Steglich, Pepin, and
  Coleman}}]{Custers:2003aa}
\bibinfo{author}{\bibfnamefont{J.}~\bibnamefont{Custers}} \bibnamefont{\textit{ et~al.}},
  \bibinfo{journal}{Nature} \textbf{\bibinfo{volume}{424}},
  \bibinfo{pages}{524} (\bibinfo{year}{2003}).

\bibitem[{\citenamefont{Yeh et~al.}(2002)\citenamefont{Yeh, Soh, Brooke,
  Aeppli, Rosenbaum, and Hayden}}]{Yeh:2002aa}
\bibinfo{author}{\bibfnamefont{A.}~\bibnamefont{Yeh}} \bibnamefont{\textit{ et~al.}}, \bibinfo{journal}{Nature}
  \textbf{\bibinfo{volume}{419}}, \bibinfo{pages}{459 } (\bibinfo{year}{2002}).

\bibitem[{\citenamefont{Greiner et~al.}(2002)\citenamefont{Greiner, Mandel,
  Esslinger, H{\~A}¤nsch, and Bloch}}]{Greiner:2002aa}
\bibinfo{author}{\bibfnamefont{M.}~\bibnamefont{Greiner}} \bibnamefont{\textit{ et~al.}},
  \bibinfo{journal}{Nature} \textbf{\bibinfo{volume}{415}}, \bibinfo{pages}{39
  } (\bibinfo{year}{2002}).

\bibitem[{\citenamefont{Canfield}(2008)}]{Canfield:2008aa}
\bibinfo{author}{\bibfnamefont{P.~C.} \bibnamefont{Canfield}},
  \bibinfo{journal}{Nature Phys.} \textbf{\bibinfo{volume}{4}},
  \bibinfo{pages}{167} (\bibinfo{year}{2008});
\bibinfo{author}{\bibfnamefont{P.}~\bibnamefont{Gegenwart}},
  \bibinfo{author}{\bibfnamefont{Q.}~\bibnamefont{Si}}, \bibnamefont{and}
  \bibinfo{author}{\bibfnamefont{F.}~\bibnamefont{Steglich}},
  \bibinfo{journal}{\textit{ibid.}} \textbf{\bibinfo{volume}{4}},
  \bibinfo{pages}{186 } (\bibinfo{year}{2008}).

\bibitem[{\citenamefont{Osterloh et~al.}(2002)\citenamefont{Osterloh, Amico,
  Falci, and Fazio}}]{Osterloh:2002aa}
\bibinfo{author}{\bibfnamefont{A.}~\bibnamefont{Osterloh}} \bibnamefont{\textit{ et~al.}},  \bibinfo{journal}{Nature} \textbf{\bibinfo{volume}{416}},
  \bibinfo{pages}{608} (\bibinfo{year}{2002}).

\bibitem[{\citenamefont{Editorial}(2008)}]{QPTfocus}
\bibinfo{author}{\bibnamefont{Editorial}}, \bibinfo{journal}{Nature Phys.}
  \textbf{\bibinfo{volume}{4}}, \bibinfo{pages}{157 } (\bibinfo{year}{2008});
\bibinfo{author}{\bibfnamefont{T.}~\bibnamefont{Giamarchi}}\bibnamefont{\textit{ et~al.}},
  \bibinfo{journal}{\textit{ibid.}} \textbf{\bibinfo{volume}{4}},
  \bibinfo{pages}{198 } (\bibinfo{year}{2008});
\bibinfo{author}{\bibfnamefont{S.}~\bibnamefont{Sachdev}},
  \bibinfo{journal}{\textit{ibid.}} \textbf{\bibinfo{volume}{4}},
  \bibinfo{pages}{173 } (\bibinfo{year}{2008}).

\bibitem[{\citenamefont{Buchler et~al.}(2007)\citenamefont{Buchler, Micheli,
  and Zoller}}]{Buchler:2007aa}
\bibinfo{author}{\bibfnamefont{H.~P.} \bibnamefont{Buchler}}\bibnamefont{\textit{ et~al.}},
  \bibinfo{journal}{Nature Phys.} \textbf{\bibinfo{volume}{3}},
  \bibinfo{pages}{726 } (\bibinfo{year}{2007});
\bibinfo{author}{\bibfnamefont{C.}~\bibnamefont{D'Cruz}} \bibnamefont{and}
  \bibinfo{author}{\bibfnamefont{J.~K.} \bibnamefont{Pachos}},
  \bibinfo{journal}{Phys. Rev. A} \textbf{\bibinfo{volume}{72}},
  \bibinfo{eid}{043608} (\bibinfo{year}{2005});
\bibinfo{author}{\bibfnamefont{K.~A.} \bibnamefont{Penson}},
  \bibinfo{author}{\bibfnamefont{J.~M.} \bibnamefont{Debierre}},
  \bibnamefont{and} \bibinfo{author}{\bibfnamefont{L.}~\bibnamefont{Turban}},
  \bibinfo{journal}{Phys. Rev. B} \textbf{\bibinfo{volume}{37}},
  \bibinfo{pages}{7884} (\bibinfo{year}{1988});\bibinfo{author}{\bibfnamefont{J.~C.} \bibnamefont{Angl\`es~d\char39{}Auriac}}
  \bibnamefont{and} \bibinfo{author}{\bibfnamefont{F.}~\bibnamefont{Igl\'oi}},
  \bibinfo{journal}{Phys. Rev. E} \textbf{\bibinfo{volume}{58}},
  \bibinfo{pages}{241} (\bibinfo{year}{1998});
\bibinfo{author}{\bibfnamefont{P.}~\bibnamefont{Suranyi}},
  \bibinfo{journal}{Phys. Rev. Lett.} \textbf{\bibinfo{volume}{37}},
  \bibinfo{pages}{725} (\bibinfo{year}{1976});
\bibinfo{author}{\bibfnamefont{J.~K.} \bibnamefont{Pachos}} \bibnamefont{and}
  \bibinfo{author}{\bibfnamefont{E.}~\bibnamefont{Rico}},
  \bibinfo{journal}{Phys. Rev. A} \textbf{\bibinfo{volume}{70}},
  \bibinfo{eid}{053620} (\bibinfo{year}{2004}).
  
  
\bibitem[{\citenamefont{Feynman}(1982)}]{Feynman:1982aa}
\bibinfo{author}{\bibfnamefont{R.~P.} \bibnamefont{Feynman}},
  \bibinfo{journal}{Int. J. Theor. Phys.} \textbf{\bibinfo{volume}{21}},
  \bibinfo{pages}{467} (\bibinfo{year}{1982}).

\bibitem[{\citenamefont{Lloyd}(1996)}]{5690}
\bibinfo{author}{\bibfnamefont{S.}~\bibnamefont{Lloyd}},
  \bibinfo{journal}{Science} \textbf{\bibinfo{volume}{273}},
  \bibinfo{pages}{1073} (\bibinfo{year}{1996}).

\bibitem[{\citenamefont{Tseng et~al.}(1999)\citenamefont{Tseng, Somaroo, Sharf,
  Knill, Laflamme, Havel, and Cory}}]{Tseng:1999aa}
\bibinfo{author}{\bibfnamefont{C.~H.} \bibnamefont{Tseng}}\bibnamefont{\textit{ et~al.}},
  \bibinfo{journal}{Phys. Rev. A} \textbf{\bibinfo{volume}{61}},
  \bibinfo{pages}{012302} (\bibinfo{year}{1999}).

\bibitem[{\citenamefont{Peng et~al.}(2005)\citenamefont{Peng, Du, and
  Suter}}]{Peng:2005aa}
\bibinfo{author}{\bibfnamefont{X.}~\bibnamefont{Peng}}\bibnamefont{\textit{ et~al.}},
  \bibinfo{journal}{Phys. Rev. A} \textbf{\bibinfo{volume}{71}},
  \bibinfo{eid}{012307} (\bibinfo{year}{2005});
\bibinfo{author}{\bibfnamefont{J.}~\bibnamefont{Zhang}}\bibnamefont{\textit{ et~al.}},  \bibinfo{journal}{Phys. Rev. Lett.} \textbf{\bibinfo{volume}{100}},
  \bibinfo{eid}{100501} (\bibinfo{year}{2008}).

\bibitem[{\citenamefont{Friedenauer et~al.}(2008)\citenamefont{Friedenauer,
  Schmitz, Gluechert, Porras, and Schaetz}}]{Friedenauer:2008aa}
\bibinfo{author}{\bibfnamefont{A.}~\bibnamefont{Friedenauer}}\bibnamefont{\textit{ et~al.}},  \bibinfo{journal}{Nature Phys.} \textbf{\bibinfo{volume}{4}},
  \bibinfo{pages}{757 } (\bibinfo{year}{2008}).

  \bibitem[{\citenamefont{Wannier}(1950)}]{Wannier:1950}
  \bibinfo{author}{\bibfnamefont{G.~H.} \bibnamefont{Wannier}},
  \bibinfo{journal}{Phys. Rev.} \textbf{\bibinfo{volume}{79}},
  \bibinfo{pages}{357} (\bibinfo{year}{1950}). 

  \bibitem[{\citenamefont{Pachos and Plenio}(2004)}]{Pachos:2004ab}
\bibinfo{author}{\bibfnamefont{J.~K.} \bibnamefont{Pachos}} \bibnamefont{and}
  \bibinfo{author}{\bibfnamefont{M.~B.} \bibnamefont{Plenio}},
  \bibinfo{journal}{Phys. Rev. Lett.} \textbf{\bibinfo{volume}{93}},
  \bibinfo{eid}{056402} (\bibinfo{year}{2004}).
  
\bibitem[{\citenamefont{Francesco et~al.}(1997)\citenamefont{Francesco,
  Mathieu, and Senechal}}]{Francesco:1997wj}
\bibinfo{author}{\bibfnamefont{P.~D.} \bibnamefont{Francesco}} \bibnamefont{\textit{ et~al.}},
  \emph{\bibinfo{title}{Conformal Field Theory}} (\bibinfo{publisher}{Springer,
  New York}, \bibinfo{year}{1997}).
  
\bibitem[{\citenamefont{Ong and Cava}(2004)}]{Ong:2004fe}
\bibinfo{author}{\bibfnamefont{N.~P.} \bibnamefont{Ong}} \bibnamefont{and}
  \bibinfo{author}{\bibfnamefont{R.~J.} \bibnamefont{Cava}},
  \bibinfo{journal}{Science} \textbf{\bibinfo{volume}{305}}, \bibinfo{pages}{52
  } (\bibinfo{year}{2004}); 
\bibinfo{author}{\bibfnamefont{S.} \bibnamefont{Nakatsuji}} \bibnamefont{\textit{ et~al.}},
  \bibinfo{journal}{Science} \textbf{\bibinfo{volume}{309}},
  \bibinfo{pages}{1697} (\bibinfo{year}{2005}); \bibinfo{author}{\bibfnamefont{G.}~\bibnamefont{Aeppli}} \bibnamefont{and}
  \bibinfo{author}{\bibfnamefont{P.}~\bibnamefont{Chandra}},
  \bibinfo{journal}{Science} \textbf{\bibinfo{volume}{275}},
  \bibinfo{pages}{177} (\bibinfo{year}{1997});
\bibinfo{author}{\bibfnamefont{M.}~\bibnamefont{Harris}},
  \bibinfo{journal}{Nature} \textbf{\bibinfo{volume}{456}},
  \bibinfo{pages}{886} (\bibinfo{year}{2008}).
  
\bibitem[{\citenamefont{Sanpera et~al.}(2001)\citenamefont{Sanpera, Bru\ss{},
  and Lewenstein}}]{PhysRevA.63.050301}
\bibinfo{author}{\bibfnamefont{A.}~\bibnamefont{Sanpera}}\bibnamefont{\textit{ et~al.}},
  \bibinfo{journal}{Phys. Rev. A} \textbf{\bibinfo{volume}{63}},
  \bibinfo{pages}{050301} (\bibinfo{year}{2001}); \bibinfo{author}{\bibfnamefont{M.}~\bibnamefont{Bourennane}}\bibnamefont{\textit{ et~al.}},
  \bibinfo{journal}{Phys. Rev. Lett.} \textbf{\bibinfo{volume}{92}},
  \bibinfo{eid}{087902} (\bibinfo{year}{2004}).
  
\bibitem[{\citenamefont{Penson et~al.}(1982)\citenamefont{Penson, Jullien, and
  Pfeuty}}]{PhysRevB.26.6334}
\bibinfo{author}{\bibfnamefont{K.~A.} \bibnamefont{Penson}}\bibnamefont{\textit{ et~al.}},
  \bibinfo{journal}{Phys. Rev. B} \textbf{\bibinfo{volume}{26}},
  \bibinfo{pages}{6334} (\bibinfo{year}{1982}).

\bibitem[{\citenamefont{D\"ur et~al.}(2000)\citenamefont{D\"ur, Vidal, and
  Cirac}}]{Dur:2000aa}
\bibinfo{author}{\bibfnamefont{W.}~\bibnamefont{D\"ur}}\bibnamefont{\textit{ et~al.}},
  \bibinfo{journal}{Phys. Rev. A} \textbf{\bibinfo{volume}{62}},
  \bibinfo{pages}{062314} (\bibinfo{year}{2000}).

\bibitem[{\citenamefont{Linden et~al.}(1999)\citenamefont{Linden, Herve,
  Carbajo, and Freeman}}]{Linden:1999ab}
\bibinfo{author}{\bibfnamefont{N.}~\bibnamefont{Linden}}\bibnamefont{\textit{ et~al.}},
  \bibinfo{journal}{Chem. Phys. Lett.} \textbf{\bibinfo{volume}{305}},
  \bibinfo{pages}{28} (\bibinfo{year}{1999}).

\bibitem[{\citenamefont{Messiah}(1976)}]{Messiah:1976aa}
\bibinfo{author}{\bibfnamefont{A.}~\bibnamefont{Messiah}},
  \emph{\bibinfo{title}{Quantum Mechanics}} (\bibinfo{publisher}{Wiley, New
  York}, \bibinfo{year}{1976}).

\bibitem[{\citenamefont{Steffen et~al.}(2003)\citenamefont{Steffen, van Dam,
  Hogg, Breyta, and Chuang}}]{Steffen:2003aa}
\bibinfo{author}{\bibfnamefont{M.}~\bibnamefont{Steffen}}\bibnamefont{\textit{ et~al.}},
  \bibinfo{journal}{Phys. Rev. Lett.} \textbf{\bibinfo{volume}{90}},
  \bibinfo{pages}{067903} (\bibinfo{year}{2003}).

\bibitem[{\citenamefont{Vandersypen et~al.}(2001)\citenamefont{Vandersypen,
  Steffen, Breyta, Yannoni, Sherwood, and Chuang}}]{Vandersypen:2001aa}
\bibinfo{author}{\bibfnamefont{L.~M.~K.} \bibnamefont{Vandersypen}}\bibnamefont{\textit{ et~al.}}, \bibinfo{journal}{Nature}
  \textbf{\bibinfo{volume}{414}}, \bibinfo{pages}{883 } (\bibinfo{year}{2001}).

\bibitem[{\citenamefont{Cory et~al.}(1998{\natexlab{a}})\citenamefont{Cory,
  Price, and Havel}}]{Cory:1998aa}
\bibinfo{author}{\bibfnamefont{D.~G.} \bibnamefont{Cory}}\bibnamefont{\textit{ et~al.}},  \bibinfo{journal}{Physica D} \textbf{\bibinfo{volume}{120}},
  \bibinfo{pages}{82} (\bibinfo{year}{1998}{\natexlab{a}}).

\bibitem[{\citenamefont{Cory et~al.}(1998{\natexlab{b}})\citenamefont{Cory,
  Price, Maas, Knill, Laflamme, Zurek, Havel, and Somaroo}}]{Cory:1998ab}
\bibinfo{author}{\bibfnamefont{D.~G.} \bibnamefont{Cory}}\bibnamefont{\textit{ et~al.}}, 
  \bibinfo{journal}{Phys. Rev. Lett.} \textbf{\bibinfo{volume}{81}},
  \bibinfo{pages}{2152} (\bibinfo{year}{1998}{\natexlab{b}});
\bibinfo{author}{\bibfnamefont{G.}~\bibnamefont{Teklemariam}}\bibnamefont{\textit{ et~al.}},
  \bibinfo{journal}{\textit{ibid.}} \textbf{\bibinfo{volume}{86}},
  \bibinfo{pages}{5845} (\bibinfo{year}{2001});\bibinfo{author}{\bibfnamefont{M.~A.} \bibnamefont{Nielsen}}\bibnamefont{\textit{ et~al.}},
  \bibinfo{journal}{Nature} \textbf{\bibinfo{volume}{396}}, \bibinfo{pages}{52
  } (\bibinfo{year}{1998}).

\bibitem[{\citenamefont{Chuang et~al.}(1998)\citenamefont{Chuang, Gershenfeld.,
  Kubinec, and Leung}}]{Chuang:1998aa}
\bibinfo{author}{\bibfnamefont{I.~L.} \bibnamefont{Chuang}}\bibnamefont{\textit{ et~al.}},
  \bibinfo{journal}{Proc. R. Soc. Lond. A} \textbf{\bibinfo{volume}{454}},
  \bibinfo{pages}{447} (\bibinfo{year}{1998}).

\bibitem[{\citenamefont{Fortunato et~al.}(2002)\citenamefont{Fortunato, Pravia,
  Boulant, Teklemariam, Havel, and Cory}}]{Fortunato:2002aa}
\bibinfo{author}{\bibfnamefont{E.~M.} \bibnamefont{Fortunato}}\bibnamefont{\textit{ et~al.}},  \bibinfo{journal}{J. Chem. Phys.} \textbf{\bibinfo{volume}{116}},
  \bibinfo{pages}{7599} (\bibinfo{year}{2002}).


\bibitem[{\citenamefont{de~Oliveira et~al.}(2006)\citenamefont{de~Oliveira,
  Rigolin, de~Oliveira, and Miranda}}]{oliveira:170401}
\bibinfo{author}{\bibfnamefont{T.~R.} \bibnamefont{de~Oliveira}}\bibnamefont{\textit{ et~al.}},  \bibinfo{journal}{Phys. Rev. Lett.} \textbf{\bibinfo{volume}{97}},
  \bibinfo{eid}{170401} (\bibinfo{year}{2006}).

\end{thebibliography}

\end{document}